\documentclass[aps,pre,twocolumn,superscriptaddress,floatfix,nofootinbib,longbibliography]{revtex4-2}
\usepackage{amsmath,amssymb,bm}
\usepackage{graphicx}
\usepackage{booktabs}
\usepackage{xcolor}
\usepackage[colorlinks=true,linkcolor=blue!60!black,citecolor=blue!60!black,urlcolor=blue!60!black]{hyperref}

\newcommand{\E}{\mathbb{E}}
\newcommand{\Var}{\operatorname{Var}}
\newcommand{\sw}{\sigma_w^2}
\newcommand{\sbb}{\sigma_b^2}
\newcommand{\qs}{q^{*}}
\newcommand{\SM}{SM}
\makeatletter
\newcommand{\lastheading}{\par\vspace*{0.65cm}\@nobreaktrue}
\makeatother

\begin{document}

\title{Finite-width shift of the edge of chaos in deep neural networks}

\author{Arn Caesar}
\email[Contact author: ]{arn.caesar@me.com}
\affiliation{Blackett Laboratory and Centre for Complexity Science, Imperial College London, London SW7 2AZ, United Kingdom}

\date{\today}

\begin{abstract}
Deep neural networks train best when they are initialised at the edge of chaos, the critical line in the plane of weight and bias variances on which the distance between two inputs neither grows nor shrinks with depth. The line is known for infinitely wide networks; at finite width it moves, and how it moves has been disputed. We treat the logarithm of the distance between two inputs through a network of width $N$ as a random walk and compute its drift and variance to order $1/N$. The edge of chaos moves to a larger weight variance, by $\Delta\sw=-a_1/(N\,\partial_{\sw}\ln\chi_1)$, about $8/N$ for $\tanh$ networks at a bias variance of $0.05$, confirmed by simulations of the exact finite-width recursion. The shift describes a typical network; the average over many networks is dominated by rare networks in which the distance grows, and its critical point moves the other way, which explains why calculations and simulations have disagreed. Backpropagation follows the same random walk, so the shift applies to gradients unchanged. Along the edge of chaos, the information about the similarity of two inputs is constant until the compression over the depth exceeds about one nat, which fixes where to sit on the line. The results give an initialisation rule, Crinit, that places a network on its finite-width edge of chaos without a search. On MNIST and Fashion-MNIST at width $64$ it is more accurate than the standard rules at their usual settings, by a small margin in shallow networks and a large one in deep networks, where most standard rules fail; it trains in fewer steps, and it matches or beats the standard rules with a tuned gain.
\end{abstract}

\maketitle

\section{Introduction}

A deep neural network starts as a random function. Its weights and biases are initialised with variances $\sw$ and $\sbb$, and whether it can be trained depends on this choice \cite{glorot2010,he2015,saxe2014}. Mean-field theory, which is exact at infinite width $N$, explains why \cite{poole2016,schoenholz2017}. It has two phases. In the ordered phase the distance between two inputs shrinks with depth, so the network forgets its input; in the chaotic phase the distance grows, so nearby inputs are torn apart. On the line between them, the edge of chaos, the distance is preserved: information and gradients propagate to any depth, and initialising there makes very deep networks trainable without normalisation layers or skip connections \cite{schoenholz2017,pennington2017,hayou2019,martens2021,zhang2022}. We call this line, which is known in closed form, the mean-field edge of chaos.

Real networks have finite width. Mean-field theory replaces every average over the $N$ neurons of a layer by its expectation; at finite $N$ such an average fluctuates about its expectation from layer to layer, by a relative amount of order $N^{-1/2}$. For one input the consequence is known: the logarithm of its norm performs a random walk, and the weight variance that keeps the typical norm fixed exceeds the mean-field value by order $1/N$ \cite{sussillo2014,hanin_nica2020,hanin_rolnick2018,cohen1984,newman1986,gurbuzbalaban2021,kogler2026}. For two inputs, and so for the edge of chaos itself, the consequence is disputed. Expansions of the mean overlap of two inputs find no shift of the critical point at order $1/N$ \cite{grosvenor2022,terin2026}, while simulations find the transition at a larger weight variance, by an amount that falls with width \cite{tamai2025,vock2025,ghavasieh2025a}, and maps of the boundary at fixed seed show finite-size effects \cite{dinverno2025}. Tamai \emph{et al.}\ fitted the extra decay of the mean decorrelation of two inputs by a rate $\mu/N$ and left its calculation open \cite{tamai2025}. A second question is where on the edge of chaos to sit: the distance is preserved exactly only where the network is linear \cite{hayou2019,murray2022,zhang2022}, so nonlinearity must be traded against propagation.

We answer both questions. The logarithm of the distance between two nearby inputs performs a random walk with drift $\ln\chi_1+a_1/N$, where $\chi_1$ is the mean-field growth factor of the distance and $a_1<0$ is an explicit Gaussian average of the activation function. The edge of chaos of a network of width $N$, which we call the finite-width edge of chaos, is where this drift vanishes, and it lies at a larger weight variance than the mean-field edge of chaos. It describes a typical network; the average over many networks shifts the other way, which explains the dispute. Backpropagated gradients follow the same walk. Along the edge of chaos, the information about input similarity is constant up to a compression of one nat. From these results we build an initialisation rule, Crinit, and test it in large sets of training runs.

\begin{figure*}[t]
\centering\includegraphics[width=\textwidth]{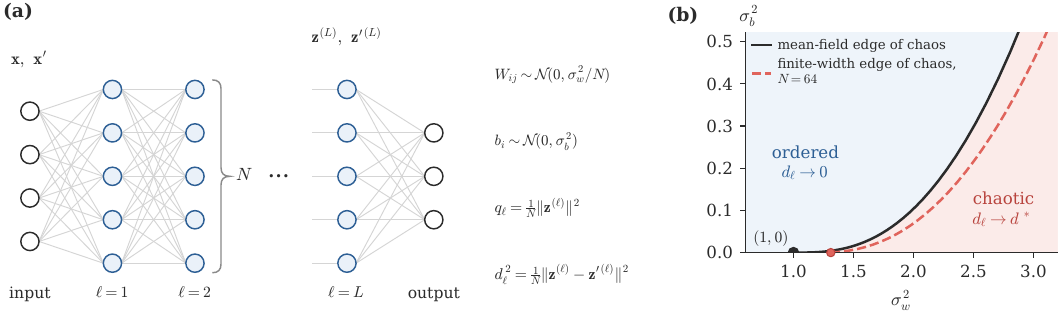}
\caption{\textbf{Deep random network and its phase diagram.} (a) A fully connected network with $L$ hidden layers of width $N$. Two inputs $\mathbf{x}$ and $\mathbf{x}'$ pass through the same network; $d_\ell$ is the distance between their pre-activation vectors $\mathbf z^{(\ell)}$ and $\mathbf z'^{(\ell)}$, divided by $\sqrt N$. (b) Phase diagram of the $\tanh$ network. On the mean-field edge of chaos (solid line) a small distance is preserved, $\chi_1=1$; below it the distance shrinks (ordered phase), above it grows to $d^*$ (chaotic phase). The finite-width edge of chaos, on which the typical distance in a network of width $N$ is preserved, lies at larger $\sw$ (dashed, $N=64$), and meets the zero-bias axis at $\sw=1.31$ (red dot; Sec.~III\,B).}
\label{fig:model}
\end{figure*}

\section{Model}

\subsection{Mean-field edge of chaos}

We consider fully connected networks with $L$ hidden layers of width $N$ [Fig.~\ref{fig:model}(a)],
\begin{equation}
 z^{(\ell+1)}_i=\sum_{j=1}^{N}W^{(\ell+1)}_{ij}\,\phi\big(z^{(\ell)}_j\big)+b^{(\ell+1)}_i ,
 \label{eq:layer}
\end{equation}
with independent weights $W_{ij}\sim\mathcal N(0,\sw/N)$ and biases $b_i\sim\mathcal N(0,\sbb)$, and an activation with $\phi(0)=0$; we use $\tanh$ in the figures and $\mathrm{erf}$ as a check. For one input, the signal size $q_\ell=\frac1N\|z^{(\ell)}\|^2$ is the mean square of the pre-activations of layer $\ell$. For two inputs $\mathbf x$ and $\mathbf x'$, $d_\ell^2=\frac1N\|z^{(\ell)}-z'^{(\ell)}\|^2$ is their squared distance. At infinite width both are the same in every network; at finite width they differ from network to network, and we then write $Q_\ell$ and $D_\ell$ for their values in one network (defined in Sec.~II\,B).

At infinite width the average over the $N$ neurons of a layer in Eq.~(\ref{eq:layer}) equals its expectation, a Gaussian average \cite{poole2016,schoenholz2017}, and the signal size follows the variance map
\begin{equation}
 q_{\ell+1}=F(q_\ell)\equiv\sw V(q_\ell)+\sbb ,
 \label{eq:vmap}
\end{equation}
where $V(q)=\E[\phi(\sqrt q\,\zeta)^2]$ is the mean square of the activation of a Gaussian pre-activation of variance $q$, with $\zeta$ a standard Gaussian. The signal settles at the steady value $\qs=F(\qs)$. The steady signal sets how nonlinear the network is: for $\tanh$, pre-activations with $\qs\ll1$ stay in the linear part of the function, $\tanh z\approx z$, so the network is nearly linear, while for $\qs$ of order one they reach its curved, saturating part. A small distance between two inputs is multiplied in every layer by
\begin{equation}
 \chi_1=\sw P(\qs),\qquad P(q)=\E\big[\phi'(\sqrt q\,\zeta)^2\big],
 \label{eq:chi}
\end{equation}
so it shrinks for $\chi_1<1$ (ordered phase) and grows for $\chi_1>1$ (chaotic phase). The mean-field edge of chaos is the line $\chi_1=1$. Each of its points has its own steady signal, and solving $\chi_1=1$ and $\qs=F(\qs)$ for the two variances, the point with steady signal $\qs$ is found at
\begin{equation}
 \sw=\frac{1}{P(\qs)},\qquad \sbb=\qs-\frac{V(\qs)}{P(\qs)} .
 \label{eq:eoc}
\end{equation}
For $\tanh$ the line starts at $(\sw,\sbb)=(1,0)$, where $\qs=0$ and the network is linear, and $\qs$ grows along it [Fig.~\ref{fig:model}(b)]. ReLU-type activations are scale free, so their edge of chaos is a single point \cite{hayou2019} (Sec.~III\,E).

Two more mean-field quantities are needed later. The first is the slope $F'=F'(\qs)$ of the variance map at its fixed point: a signal size that starts away from $\qs$ approaches it by the factor $F'$ per layer, so $1/(1-F')$ is roughly the number of layers the signal needs to settle. The second is the compression of distances that are not small. Let $C_\ell$ be the correlation of the two inputs at layer $\ell$ and $\varepsilon_\ell=1-C_\ell$ their decorrelation, which is zero for identical inputs and equals $d_\ell^2/(2\qs)$. Expanding the mean-field map of the correlation about $C=1$ to second order gives \cite{hayou2019,tamai2025}
\begin{equation}
 \varepsilon_{\ell+1}=\chi_1\varepsilon_\ell-\kappa\,\varepsilon_\ell^2,\qquad \kappa=\tfrac12\sw\qs\E[\phi''^2],
 \label{eq:cmap}
\end{equation}
where the quadratic term comes from the curvature of $\phi$. On the mean-field edge of chaos the linear term preserves $\varepsilon$ and the quadratic term lowers it, so after $\ell$ layers $\varepsilon_\ell=\varepsilon_0/(1+\kappa\varepsilon_0\ell)$, which tends to $1/(\kappa\ell)$ whatever $\varepsilon_0$ was: pairs of inputs that started with different similarities end with nearly the same one. We call this the compression of input similarities. It holds for any smooth activation, with a rate $\kappa$ that grows with the signal ($\kappa\approx2q^{*2}$ for $\tanh$ at small $\qs$); a kink, as in ReLU, changes the power (\SM, Sec.~S1 \cite{sm}).\nocite{stein1981,price1958,chernoff1981,daniely2016,jacot2022,sarmiento2026,lesser2026,sparre1954}

\subsection{Finite width: the distance as a random walk}

At finite width the signal size and the squared distance of a layer differ from network to network. Let $z_j$ and $z'_j$ be the pre-activations of the two inputs at neuron $j$ of layer $\ell$, and $\delta_j=z_j-z'_j$ their difference. For Gaussian weights, given layer $\ell$, each pre-activation of layer $\ell+1$ is a sum of $N$ independent Gaussian terms, Eq.~(\ref{eq:layer}), so it is exactly Gaussian, and the pairs $(z_i,\delta_i)$ of layer $\ell+1$ are exactly independent bivariate Gaussians with variances and covariance
\begin{align}
 Q_{\ell+1}&=\sw\tfrac1N\textstyle\sum_j\phi(z_j)^2+\sbb,\nonumber\\
 D_{\ell+1}&=\sw\tfrac1N\textstyle\sum_j\big[\phi(z_j)-\phi(z'_j)\big]^2,\nonumber\\
 K_{\ell+1}&=\sw\tfrac1N\textstyle\sum_j\phi(z_j)\big[\phi(z_j)-\phi(z'_j)\big]
 \label{eq:exact}
\end{align}
(\SM, Sec.~S2). So $Q_\ell$ and $D_\ell$ are the variances of the signal and of the difference in one network, given the layer before. The right-hand sides of Eq.~(\ref{eq:exact}) are averages over the neurons of layer $\ell$; replacing them by their expectations gives the mean-field maps of Sec.~II\,A, and keeping them lets two inputs be followed through a random network of any width by sampling these Gaussian numbers layer by layer, without weight matrices. All our simulations of the walk use this exact chain.

\begin{figure*}[t]
\centering\includegraphics[width=\textwidth]{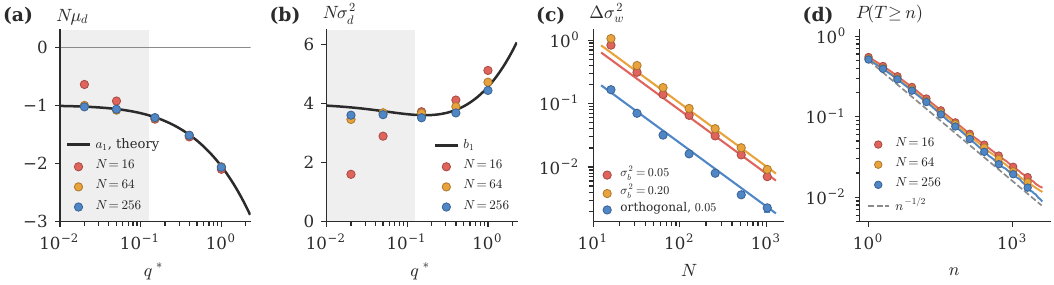}
\caption{\textbf{Drift and variance of the distance, and the finite-width edge of chaos.} (a) Drift $\mu_d$ of $\ln D_\ell$, times $N$, along the mean-field edge of chaos of the $\tanh$ network, where $\ln\chi_1=0$ and $N\mu_d=a_1$ for every $N$, against the steady signal $\qs$: Eq.~(\ref{eq:a1}) (line) and simulations of the exact chain (\ref{eq:exact}) for $N=16$, $64$ and $256$ (dots); shaded: $N(1-F')<3$ at $N=16$, where the expansion fails. (b) Long-run variance $\sigma_d^2$, times $N$: $b_1$ (line) and simulations. (c) Shift of the edge of chaos at $\sbb=0.05$ and $0.20$, Eq.~(\ref{eq:shift}) (lines), against the $\sw$ at which the measured drift changes sign (dots), for Gaussian and orthogonal weights. (d) Probability that an avalanche of the distance lasts at least $n$ layers on the finite-width edge of chaos; dashed: $n^{-1/2}$.}
\label{fig:walk}
\end{figure*}

For a small distance, $\phi(z_j)-\phi(z'_j)\approx\phi'(z_j)\delta_j$, so $D_{\ell+1}=\sw\frac1N\sum_j\phi'(z_j)^2\delta_j^2$. Dividing by $D_\ell$ and writing $\eta_j=\delta_j/\sqrt{D_\ell}$ for the difference scaled to unit variance (given the layer before, the $\eta_j$ are independent standard Gaussians), the logarithm of the squared distance changes per layer by
\begin{equation}
 s_\ell=\ln\frac{D_{\ell+1}}{D_\ell}=\ln\sw+\ln\Big[\frac1N\sum_j\phi'(z_j)^2\eta_j^2\Big].
 \label{eq:step}
\end{equation}
After $\ell$ layers, $\ln D_\ell=\ln D_0+s_0+s_1+\dots+s_{\ell-1}$: a sum of random steps, that is, a random walk, which we call $X_\ell$. At infinite width the average in the bracket equals $P(\qs)$, every step equals $\ln\sw+\ln P(\qs)=\ln\chi_1$, and the walk is a straight line, the mean-field result $D_\ell=\chi_1^\ell D_0$. At finite width the steps vary for two reasons. The average in the bracket runs over finitely many neurons, so it fluctuates about $P(\qs)$ by a relative amount of order $N^{-1/2}$. And the surroundings of the perturbation vary: the signal size $Q_\ell$ fluctuates about $\qs$, and the difference $\delta_j$ is partly aligned with the signal $z_j$, with a correlation $r_\ell=K_\ell/\sqrt{Q_\ell D_\ell}$ that fluctuates about zero. Both fluctuations are of order $N^{-1/2}$ for the same reason: $Q$ and $K$ are averages over $N$ neurons.

Write $\E[\cdot]$ for the average over networks, that is, over the random weights and biases, and keep $\frac1N\sum_j$ for averages over neurons. The drift $\mu_d=\E[s_\ell]$ is the mean step. Since the steps add, $\E[\ln D_\ell]\simeq\mu_d\ell$: the drift is the rate at which the \emph{typical} squared distance, $\exp\E[\ln D_\ell]$, grows or shrinks per layer. The typical value is close to the median, the value found in most networks. It is not the \emph{mean} $\E[D_\ell]$, because $D_\ell$ is a product of random factors. If each layer multiplied $D$ by $1.1$ or $0.9$ with equal probability, the mean would stay at one, while in most networks $D$ would fall by $1.1\times0.9=0.99$ every two layers; the mean is held up by the few networks in which most factors were $1.1$. A practitioner trains a single network, so the edge of chaos that matters is where the typical distance is preserved, $\mu_d=0$. The drift also governs the many pairs of inputs that pass through one network: averaged over the pairs, $\ln D_\ell$ grows at the rate $\mu_d$ as well, up to an offset specific to the network, which carries about two fifths of the variance of the walk (\SM, Sec.~S3.7).

\section{Results}

\subsection{Drift and variance of the distance}

The step (\ref{eq:step}) depends on the layer it is taken in through the fluctuating average in the bracket and through the surroundings $(Q_\ell,r_\ell)$. After the first few layers the statistics of the surroundings no longer depend on $\ell$. The drift then follows in three steps (\SM, Sec.~S3). First, write the average in the bracket as its mean times $1+\epsilon$, where $\epsilon$ has zero mean and a variance of order $1/N$, and expand $\ln(1+\epsilon)=\epsilon-\frac12\epsilon^2+\dots$: the logarithm of a fluctuating average is, on average, lower than the logarithm of its mean, by half the variance of $\epsilon$. Second, the mean of the bracket depends on the surroundings; write $Q_\ell=\qs+x_\ell$ and expand it to second order in the small quantities $x_\ell$ and $r_\ell$. Third, average over networks, which brings in the stationary variances $\E[x^2]$ and $\E[r^2]$ and the small shift $\E[x]$ of the mean signal size, all of order $1/N$. Collecting the terms,
\begin{equation}
 \mu_d=\ln\chi_1+\frac{a_1}{N},\qquad \sigma_d^2=\frac{b_1}{N},
 \label{eq:drift}
\end{equation}
with
\begin{align}
 a_1&=-\frac12\Big(\frac{3P_4}{P^2}-1\Big)+2\qs\frac{P'}{P}\,N\E[r^2]+\frac{P'}{P}\,N\E[x]
 \nonumber\\&\quad+\frac12\Big(\frac{P''}{P}-\frac{P'^2}{P^2}\Big)N\E[x^2],
 \label{eq:a1}
\end{align}
where $P_4=\E[\phi'^4]$, primes are derivatives in $q$, all averages are at $\qs$, and $x=Q-\qs$. The four terms of $a_1$ are the fluctuation of the average over neurons (first term), the alignment of the perturbation with the signal (second) and the fluctuation of the signal size (third and fourth). The moments $N\E[r^2]$, $N\E[x]$ and $N\E[x^2]$ are Gaussian averages of $\phi$ (\SM, Sec.~S3); the factors of $N$ make them independent of $N$. The long-run variance $\sigma_d^2$ is the variance the walk gains per layer, $\Var X_\ell\simeq\sigma_d^2\ell$, which sets how much $\ln D_\ell$ differs from network to network after $\ell$ layers; it includes the correlations between successive steps, and $b_1$ is given in the \SM. Equations~(\ref{eq:drift}) and (\ref{eq:a1}) are our main result. For the expansion to be valid, the signal size must settle within far fewer than $N$ layers, $N(1-F')\gg1$, so that the surroundings forget their past before their fluctuations add up; for $\tanh$ this means $\qs\gg1/(2N)$.

The first term of $a_1$ dominates. It is the effect of the logarithm on a fluctuating average, $\E[\ln(1+\epsilon)]\approx-\frac12\Var\epsilon$ with $\Var\epsilon=(3P_4/P^2-1)/N$, and since $P_4\ge P^2$ it is negative for every activation. Alone it gives the known single-input results: $a_1=-1$ for a linear network and $a_1=-\frac12C_\alpha$ for the leaky ReLU of slope $\alpha$, with $C_\alpha=(5-2\alpha^2+5\alpha^4)/(1+\alpha^2)^2$, which is $5$ for ReLU \cite{cohen1984,newman1986,kogler2026}. Figure~\ref{fig:walk}(a,b) compares Eq.~(\ref{eq:drift}) with simulations of the exact chain (\ref{eq:exact}) along the mean-field edge of chaos. For $N=64$ and $256$ the drift agrees with $a_1$ within one to two per cent for $\qs$ from $0.02$ to $1$, and the variance with $b_1$ within a few per cent, up to eleven per cent at the smallest $\qs$; the same holds for $\mathrm{erf}$ (\SM, Fig.~S1). At $N=16$ and small $\qs$ the expansion fails, since $N(1-F')<3$ there.

\subsection{The finite-width edge of chaos}
\label{sec:shift}

The drift vanishes when $\ln\chi_1=-a_1/N$. Since $a_1<0$, this needs $\chi_1>1$: the finite-width edge of chaos lies at a larger $\sw$ than the mean-field edge of chaos. At fixed bias it is displaced by
\begin{equation}
 \Delta\sw=\frac{-a_1}{N\,\partial_{\sw}\ln\chi_1},
 \label{eq:shift}
\end{equation}
where $\partial_{\sw}\ln\chi_1=1/\sw+(P'/P)V/(1-F')$ is how fast $\ln\chi_1$ grows with $\sw$ at fixed $\sbb$ (the steady signal moves with $\sw$ too). For $\tanh$ the coefficient is about $8$: $\Delta\sw=7.9/N$ at $\sbb=0.05$. The shift has the same origin as the example of Sec.~II\,B: a product of random factors with mean one shrinks in most networks, so $\chi_1$ must exceed one for the typical distance to be preserved. Equation~(\ref{eq:shift}) is the first order of an expansion in $1/N$ and fails near the end of the line at $(1,0)$, where $\partial_{\sw}\ln\chi_1$ vanishes: along the zero-bias axis $\ln\chi_1\approx(\sw-1)^2/3$, so the zero-bias point of the finite-width edge of chaos, found by solving $\ln\chi_1=-a_1/N$ without linearising, lies at a distance of order $N^{-1/2}$ from $(1,0)$ rather than $1/N$: $\sw=1.31$ at $N=64$ and $1.13$ at $N=256$ (all orders: $1.30$, \SM, Table~S2).

To test Eq.~(\ref{eq:shift}) we locate the finite-width edge of chaos in simulations of the exact chain as the $\sw$ at which the measured drift changes sign [Fig.~\ref{fig:walk}(c), dots; $N$ from $16$ to $1024$]. For Gaussian weights and $N\ge256$ it agrees with Eq.~(\ref{eq:shift}) within the statistical error. At smaller $N$ the measured shift at fixed bias is larger than Eq.~(\ref{eq:shift}), by $13\%$ at $N=64$ ($0.14$ instead of $0.12$ in $\sw$), by a fraction that falls as $1/N$, as the next order of the expansion predicts; the all-orders value is available numerically (\SM, Sec.~S3.6). At fixed steady signal instead of fixed bias, the all-orders correction is $2$ to $9\%$ smaller than the first-order one for $N\ge64$ (\SM, Table~S2).

Orthogonal weights test the mechanism. An orthogonal matrix, a random rotation scaled by $\sigma_w$, preserves the length of every vector it multiplies, so only the nonlinearity $\phi$ makes the signal size and the squared distance fluctuate. The mean-field edge of chaos is unchanged, because it depends only on means, but the random walk has smaller fluctuations: the theory predicts $a_1=-0.51$ instead of $-1.68$ for $\tanh$ at $\sbb=0.05$ (\SM, Sec.~S4), and the measured shift is about four times smaller [Fig.~\ref{fig:walk}(c)], consistent with Refs.~\cite{saxe2014,hu2020,day2025,combette2026}. The measured drift (\SM, Table~S1, $\qs\ge0.15$) and shift ($N=64$ to $256$) lie $13$ to $18\%$ below this theory, a discrepancy of opposite sign to the Gaussian one that we have not traced.

Backpropagated gradients perform the same walk. The gradient of the loss goes down through the same layers that the signal came up, multiplied in each by the same slopes $\phi'$ and by the transpose of the same weight matrix; a matrix and its transpose stretch random vectors by the same factors, so the squared norm of a gradient with a random direction at the output has exactly the same distribution over networks as the squared distance between two nearby inputs, at any width (\SM, Sec.~S5). Its logarithm therefore performs the walk of Eqs.~(\ref{eq:step})--(\ref{eq:a1}) with the same $a_1$ and $b_1$, and the finite-width edge of chaos is also where the typical gradient neither vanishes nor explodes. The difference from the mean-field edge of chaos is large: for $\tanh$ with $N=64$ at $\sbb=0.05$, the median squared gradient $96$ layers below the output is $0.08$ of its value at the output on the mean-field edge of chaos and $0.83$ on the finite-width edge of chaos; the gradient of a real loss, the cross-entropy of a read-out layer on MNIST images, follows the same walk within the errors (\SM, Secs.~S5.4 and S5.5).

\subsection{Typical versus mean}
\label{sec:typical}

\begin{figure}[t]
\centering\includegraphics[width=\columnwidth]{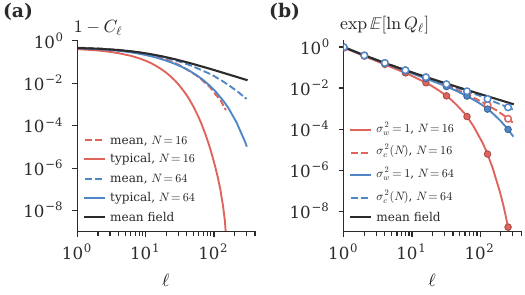}
\caption{\textbf{Typical and mean propagation differ at finite width.} (a) Decorrelation $1-C_\ell$ of two inputs ($C_0=0.5$) through $\tanh$ networks of width $16$ and $64$ on the mean-field edge of chaos ($\sw=1.95$, $\sbb=0.09$): mean over $2$--$4\times10^4$ networks (dashed, shown while its error is below $10\%$), typical value $\exp\E[\ln(1-C_\ell)]$ (solid) and the mean-field map (black). (b) Typical signal $\exp\E[\ln Q_\ell]$ of one input at $\sw=1$ (solid) and at $\sigma_c^2(N)$ (dashed), against the mean-field decay $1/(2\ell)$ (black); its recursion is in the \SM\ (Fig.~S4).}
\label{fig:typical}
\end{figure}

The walk also gives the mean distance (Fig.~\ref{fig:typical}). After many layers $X_\ell=\ln D_\ell$ is a sum of many steps, so it is approximately Gaussian with mean $\mu_d\ell$ and variance $\sigma_d^2\ell$, and the mean of $D_\ell=e^{X_\ell}$ is $\exp(\mu_d\ell+\frac12\sigma_d^2\ell)$: the mean of $e^X$ exceeds $e^{\E[X]}$ by the factor $e^{\Var X/2}$, as the mean of a log-normal variable exceeds its median. On the mean-field edge of chaos, where $\mu_d=a_1/N$, the typical distance shrinks as $\exp(a_1\ell/N)$ while the mean distance grows, at an estimated rate $(a_1+\frac12b_1)/N$ per layer; for $\tanh$ at $\sbb=0.05$ this estimate gives $a_1+\frac12b_1=+0.32$ (from the values in \SM, Table~S3). The estimate is rough, because the mean is carried by the tail of the distribution of $X_\ell$, which is not exactly Gaussian; the growth rate of the mean measured along the line, $N\mu_{\rm mean}$, lies between $+0.1$ and $+0.8$ (\SM, Table~S1). For inputs a finite distance apart the same fluctuations make the mean of $1-C_\ell$ decay faster than the mean-field map, as Tamai \emph{et al.}\ observed \cite{tamai2025}, and the typical value faster still [Fig.~\ref{fig:typical}(a)].

\begin{figure*}[t]
\centering\includegraphics[width=\textwidth]{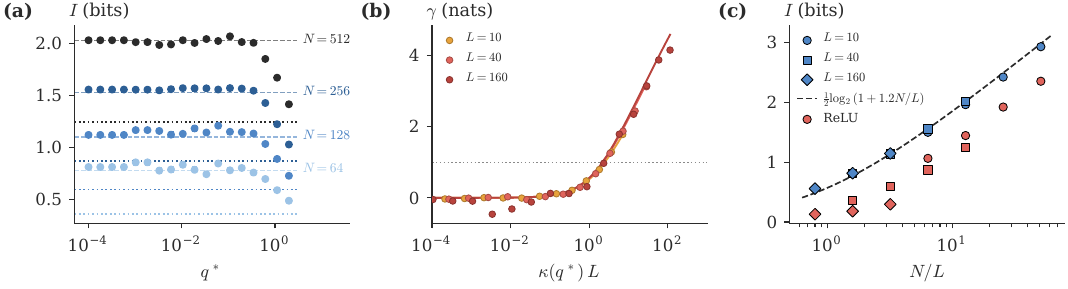}
\caption{\textbf{Information about input similarity along the edge of chaos.} (a) Information $I$ (bits per pair of inputs) transmitted through $L=40$ layers of $\tanh$ networks of width $N=64$ to $512$, against the steady signal $\qs$ on the mean-field edge of chaos (mean over $40$ networks; errors about the size of the symbols); dashed and dotted: a linear network and ReLU at He initialisation. (b) Compression $\gamma$ against $\kappa(\qs)L$ for three depths ($N=512$), with the mean-field prediction (lines); dotted: $\gamma=1$. (c) Plateau height of (a) against $N/L$ for all depths and widths (blue), ReLU (red), and $\frac12\log_2(1+1.2N/L)$ (dashed).}
\label{fig:info}
\end{figure*}

This resolves the disagreement in the literature. Refs.~\cite{grosvenor2022,terin2026} compute finite-width corrections to averages over networks as a perturbation series whose terms are of order depth over width \cite{pdlt,hanin2024}: Ref.~\cite{grosvenor2022} for the two-point function of one input across layers, Ref.~\cite{terin2026} for its dressed kernels. Both find the critical point unchanged at the order considered; Ref.~\cite{grosvenor2022} adds that this conclusion ``should however be regarded with skepticism, since the theory breaks down at the critical point itself'', and Ref.~\cite{terin2026} that higher orders may shift it. There is no contradiction. At fixed depth a displacement of the critical point of order $1/N$ is the same thing as a correction of relative order $\ell/N$ to the kernels, since $e^{c\ell/N}=1+c\ell/N+\dots$; it shows as a displaced critical point only once the depth is comparable to $N$, where the series must be resummed, which the walk does: the weight variance at which the mean distance is preserved lies below the mean-field edge of chaos, by an amount of order $1/N$, and the one at which the typical distance is preserved lies above it, by $\Delta\sw$ of Eq.~(\ref{eq:shift}). Simulations \cite{tamai2025,vock2025,ghavasieh2025a,dinverno2025} follow single networks and see the typical distance. The rate $\mu/N$ fitted by Tamai \emph{et al.}\ describes the mean of $1-C_\ell$ over networks, so it is not $a_1$, the drift of the typical value; the walk relates the two. That typical and mean growth differ is well known for products of random matrices: the typical growth rate is set by the largest Lyapunov exponent, which is self-averaging, and the growth of the mean by a generalised Lyapunov exponent that exceeds it \cite{crisanti1993}. The drift $\mu_d$ of the squared distance is twice the largest Lyapunov exponent, and Eq.~(\ref{eq:drift}) gives its $1/N$ correction. A single input shows the same split [Fig.~\ref{fig:typical}(b)]. The point $\sw=1/\phi'(0)^2$ of LeCun, Glorot and He \cite{glorot2010,he2015} keeps the mean squared norm fixed, while the typical norm shrinks by $e^{-1/N}$ per layer; the typical norm is kept fixed by $\sw=\sigma_c^2(N)=\frac1{\phi'(0)^2}\frac N2e^{-\psi(N/2)}\approx\frac{1}{\phi'(0)^2}(1+\frac1N)$, with $\psi$ the digamma function. This typical-preserving variance has been derived several times \cite{cohen1984,newman1986,sussillo2014,gurbuzbalaban2021,kogler2026} but has not entered practice; it is the single-input special case of Eq.~(\ref{eq:shift}).

\subsection{Where to sit on the edge of chaos}

The edge of chaos fixes one combination of $\sw$ and $\sbb$ and leaves the steady signal $\qs$ free. A larger $\qs$ makes the network more nonlinear, which it needs to represent nonlinear functions, but it also compresses input similarities faster, since the compression rate $\kappa(\qs)$ of Sec.~II\,A grows with the signal. To find the cost of each choice, we measured how much information about the similarity of two inputs survives $L$ layers of a finite $\tanh$ network: $128$ pairs of inputs with decorrelations $\varepsilon_0$ spread evenly on a logarithmic scale over $[10^{-3},1]$ are sent through the same network, and the output decorrelation $\varepsilon_L$ of every pair is recorded. In a network that preserved similarities perfectly, $\ln\varepsilon_L$ would be a fixed function of $\ln\varepsilon_0$; in a real network the points $(\ln\varepsilon_0,\ln\varepsilon_L)$ scatter about a curve. We fit the curve by a cubic polynomial $f$, which plays the role of the signal, treat the scatter about it as noise, and quote the capacity of a Gaussian channel with this signal and noise \cite{cover2006}, $I=\tfrac12\log_2[1+\Var f/\Var(\ln\varepsilon_L-f)]$, in bits per pair of inputs. This is a proxy for the information about input similarity, not a bound on it, since it treats the scatter as Gaussian noise; polynomials of degree two to five change $I$ by less than the statistical error (\SM, Sec.~S7). Two things lower $I$: compression, which flattens the curve so that different input similarities give nearly the same output similarity, and finite-width noise, which scatters the points about the curve. Figure~\ref{fig:info} gives three results. (a) Along the mean-field edge of chaos, $I$ does not depend on $\qs$ up to $\qs\approx0.1$--$0.3$ at $L=40$, where it equals that of a linear network of the same width: small signals are not lost in the finite-width noise. At larger $\qs$ the information falls. (b) The fall is caused by compression, which we measure by how much the range of similarities shrinks on the logarithmic scale between input and output, $\gamma=\ln(\varepsilon_0^{\max}/\varepsilon_0^{\min})-\ln(\varepsilon_L^{\max}/\varepsilon_L^{\min})$, in nats ($\gamma=1$: shrunk by a factor $e$). By the compression law of Sec.~II\,A, $\gamma$ depends on $\qs$ and $L$ only through $\kappa(\qs)L$, and the measured $\gamma$ of all depths falls on the mean-field curve, within the scatter at $L=160$. The information starts to fall where $\gamma\approx1$, at $\kappa L\approx2$; we call $\gamma=1$ the compression budget, the compression a network can apply over its depth before it loses information about input similarity. (c) The height of the plateau, the value of $I$ where it does not depend on $\qs$, depends on $N$ and $L$ only through $N/L$ and is fitted by $\frac12\log_2(1+cN/L)$ with $c=1.2$: the capacity of a Gaussian channel with signal-to-noise ratio $cN/L$, as expected, since the walk of Eq.~(\ref{eq:drift}) adds a variance of order $1/N$ to $\ln D$ at every layer, so that after $L$ layers the noise in $\ln\varepsilon_L$ is of order $L/N$ \cite{jakub2024,combette2026}. ReLU, whose edge of chaos is a single point at which all inputs are driven together \cite{hayou2019}, transmits $1.3$ to $4.6$ times fewer bits than $\tanh$ on its edge of chaos.

\begin{figure*}[t]
\centering\includegraphics[width=\textwidth]{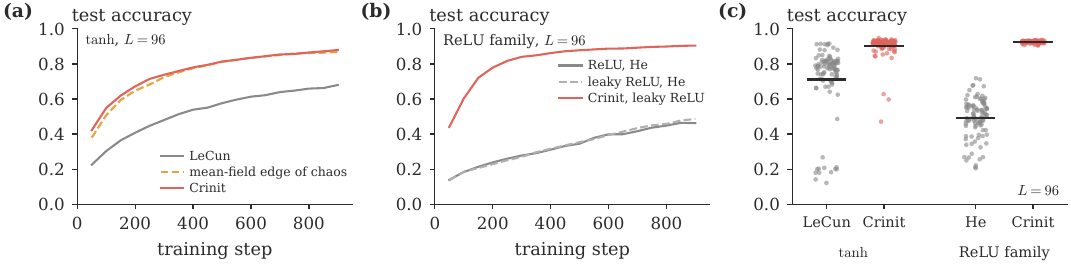}
\caption{\textbf{Training from the finite-width edge of chaos} (MNIST, $N=64$, $L=96$, Gaussian weights). (a) Mean test accuracy of one hundred $\tanh$ networks against the training step, for LeCun's rule, the mean-field edge of chaos at the Crinit operating point, and Crinit. (b) The same for ReLU and the leaky ReLU (slope $0.01$) at He's rule, and Crinit's leaky ReLU (mean over the $79$ networks that did not diverge at its pilot's learning rate). (c) Final test accuracy of each network that did not diverge; bars: means.}
\label{fig:mnist}
\end{figure*}

The operating point is therefore the largest $\qs$ within the budget,
\begin{equation}
 \gamma(\qs,L)=1\qquad(\kappa L\approx2),
 \label{eq:sweet}
\end{equation}
which gives $\qs=0.31$ for $L=32$, $\qs=0.14$ for $L=96$, and $\qs\propto L^{-1/2}$ at large depth, since $\kappa\propto q^{*2}$. The value $\gamma=1$ is measured, not derived; training scans find the best accuracy for $\gamma$ between $0.5$ and $1$ (\SM, Secs.~S8 and S9). Equation~(\ref{eq:sweet}) has the form of the depth rules of Refs.~\cite{hayou2019,martens2021,zhang2022}; the measurement supplies the value of the budget.

\subsection{Crinit: an initialisation for finite width}

Crinit (critical initialisation) combines the two results; we provide it as a software package \cite{code}. Given the activation, the depth $L$ and the width $N$, it takes $\qs$ from Eq.~(\ref{eq:sweet}) and places the network on the finite-width edge of chaos at that signal: $\sw=e^{-a_1/N}/P(\qs)$, the mean-field value of Eq.~(\ref{eq:eoc}) with the correction of Eq.~(\ref{eq:drift}), and $\sbb=\qs-\sw V(\qs)$, the bias that makes $\qs$ the steady signal. The finite-width edge of chaos with $\sbb\ge0$ begins at the zero-bias point of Sec.~III\,B, $\sw=1.31$ for $\tanh$ at $N=64$; when the $\qs$ of Eq.~(\ref{eq:sweet}) would need a negative bias variance, Crinit uses this point instead, which at $N=64$ happens for $L=96$. Crinit also sets the first layer: its weights have variance $(\qs-\sbb)/d_{\rm in}$, with $d_{\rm in}$ the dimension of the input and the inputs standardised to unit variance, so that the first hidden layer already carries the steady signal $\qs$, whereas the standard rules use $\sw/d_{\rm in}$ there and need several layers for the signal to settle. For ReLU-type activations, whose edge of chaos is a point, the compression is set by the slope $\alpha$ of a leaky ReLU \cite{zhang2022}, which Crinit chooses so that $\gamma=1$, with $\sw=\frac2{1+\alpha^2}e^{C_\alpha/(2N)}$ and zero bias; since these activations are scale free, the signal has no steady value, and Crinit gives the first hidden layer unit signal, with first-layer variance $1/d_{\rm in}$. Crinit has one parameter, the compression $\gamma$, with default $1$.

We trained networks of width $N=64$ and depth $L=32$ and $96$ on MNIST \cite{lecun1998} with stochastic gradient descent with momentum for two epochs, without normalisation or skip connections, so that the initialisation alone decides whether a network trains. The aspect ratios $L/N=\frac12$ and $\frac32$ set the noise that the walk accumulates over the depth [Fig.~\ref{fig:info}(c)], so the deeper networks are the harder test. For each rule and depth the learning rate was chosen from three values by a pilot, and then one hundred independent networks were trained (\SM, Sec.~S8); Table~\ref{tab:mnist} and Fig.~\ref{fig:mnist} give the results.

\begin{table}[t]
\caption{\textbf{MNIST test accuracy after two epochs, and training steps to reach $85\%$ training accuracy} ($N=64$; each rule at the learning rate chosen by its pilot). Each entry is a mean over one hundred networks (steps rounded to whole numbers). The number in parentheses is the standard error of that mean (the spread of the networks divided by the square root of their number) in units of the last digit: $0.903(6)$ means $0.903\pm0.006$ (steps: \SM, Table~S6). Bold: the best value in each group and any value statistically equal to it (a difference below twice the standard error of the difference). Superscripts: the number of networks in the mean when not all (those that diverged or never reached $85\%$ are left out). Typical-preserving: $\sw=\sigma_c^2(N)$; leaky ReLU: slope $0.01$; mean-field edge of chaos: the Crinit point of its group without the $1/N$ correction.}
\label{tab:mnist}
\footnotesize\setlength{\tabcolsep}{2pt}
\begin{ruledtabular}
\begin{tabular}{lcccc}
 & \multicolumn{2}{c}{$L=32$} & \multicolumn{2}{c}{$L=96$}\\
initialisation & accuracy & steps & accuracy & steps\\
\hline
\multicolumn{5}{l}{\textit{$\tanh$, Gaussian weights}}\\
PyTorch ($\sw=\frac13$) & 0.1135 & -- & 0.1135 & --\\
LeCun ($\sw=1$) & 0.9438(5) & 257 & 0.713(21) & 690$^{10}$\\
He ($\sw=2$) & 0.9392(5) & 234 & 0.567(16) & --\\
typical-preserving & 0.9448(5) & 247 & 0.744(19) & 598$^{18}$\\
mean-field edge of chaos & \textbf{0.9513(4)} & 164 & \textbf{0.893(10)} & \textbf{504}$^{88}$\\
\cline{1-5}
\multicolumn{1}{|l}{Crinit} & \textbf{0.9517(4)} & \textbf{159} & \textbf{0.903(6)} & \multicolumn{1}{c|}{\textbf{497}$^{91}$}\\
\cline{1-5}
\multicolumn{5}{l}{\textit{$\tanh$, orthogonal weights}}\\
orthogonal ($\sw=1$) & 0.9538(4) & 168 & 0.9428(6) & 197\\
mean-field edge of chaos & \textbf{0.9581(4)} & \textbf{96} & \textbf{0.9545(5)} & \textbf{112}\\
\cline{1-5}
\multicolumn{1}{|l}{Crinit, orthogonal} & \textbf{0.9581(3)} & \textbf{95} & \textbf{0.9553(4)} & \multicolumn{1}{c|}{\textbf{111}}\\
\cline{1-5}
\multicolumn{5}{l}{\textit{ReLU family, Gaussian weights}}\\
ReLU, He ($\sw=2$) & 0.9099(8) & 573 & 0.491(11) & --\\
ReLU, typical-preserving & 0.9203(7) & 495 & 0.480(13) & --\\
leaky ReLU, He & 0.9114(7) & 553 & 0.505(11) & --\\
\cline{1-5}
\multicolumn{1}{|l}{Crinit, leaky ReLU} & \textbf{0.9404(4)} & \textbf{211} & \textbf{0.9250(8)}$^{79}$ & \multicolumn{1}{c|}{\textbf{333}$^{79}$}\\
\cline{1-5}
\end{tabular}

\end{ruledtabular}
\end{table}

At $L=32$ every rule that trains reaches a similar accuracy. Crinit's margin over the standard rules is small but resolved, at eight or more standard errors, and it needs $32$ to $63\%$ fewer steps than they do. Against the mean-field edge of chaos at the same operating point its accuracy is the same within the error, $0.9517(4)$ against $0.9513(4)$, and it is faster, $159(1)$ against $164(2)$ steps; a larger paired run confirms both (\SM, Sec.~S8). At $L=96$ the rules separate. LeCun's rule, at $\qs=0$, where the typical signal collapses as $e^{-\ell/N}$ [Fig.~\ref{fig:typical}(b)], reaches $0.713(21)$, and only $10$ of its $100$ networks reach $85\%$ training accuracy; the typical-preserving variance $\sigma_c^2(N)$ raises this to $0.744(19)$ and $18$, within the error; Crinit reaches $0.903(6)$ (median $0.919$), with $91$. In the ReLU family only Crinit's leaky ReLU trains, to $0.9250(8)$ over the $79$ networks that did not diverge at the learning rate chosen by its pilot, and to $0.9112(7)$ over all hundred at a third of that rate. Crinit is ahead of the standard rules at every step, and its gain is not carried by a few networks: $84$ of its $100$ networks lie within $0.03$ of their median, against $35$ of LeCun's (Fig.~\ref{fig:mnist}). Most of Crinit's gain over the standard rules therefore comes from the operating point and the first layer, not from the $1/N$ correction: at $N=64$ the correction moves $\sw$ by only $0.03$ at $L=32$ and $0.07$ at $L=96$, and at $L=96$ the mean-field edge of chaos reaches $0.893(10)$, a difference of $0.009(12)$ from Crinit, again within the error. With orthogonal weights the correction is smaller still ($0.006$ in $\sw$ at $L=32$), and the two are equal in accuracy and speed at both depths.

At width $N=32$, where the correction is twice as large (\SM, Sec.~S8.3), the picture is the same: at $L=32$ Crinit and the mean-field edge of chaos reach $0.9332(6)$ and $0.9326(6)$, and Crinit needs $11\%$ fewer steps, $254(4)$ against $286(7)$. At $L=96$, an aspect ratio of $3$, both are poor ($0.660(13)$ and $0.684(15)$; LeCun $0.307(12)$), because the zero-bias point compresses twice the budget while the mean-field point contracts the typical distance by $e^{-3.6}$; only Crinit's leaky ReLU, which has no signal size to keep, trains well there ($0.894(2)$ over the $78$ networks that did not diverge; He $0.423(12)$).

On Fashion-MNIST \cite{xiao2017} we also tuned, on a validation set, the gain of the standard rules and the compression $\gamma$ of Crinit (\SM, Sec.~S9). Over twelve comparisons, Crinit beats every standard rule at its usual variance, by up to $0.355$ at $L=96$, beats the best tuned gain in eight and is within the error in four, and is the fastest to $80\%$ training accuracy in every group. The gain search finds the finite-width edge of chaos, its best values bracketing $\sw=1.31$ rather than $1$, but not Crinit's smaller first layer, which takes larger relative steps under gradient descent and is worth up to two per cent at $L=96$.

\section{Discussion}

The edge of chaos of a network of width $N$ lies a computable distance from the mean-field edge of chaos, about $8/N$ in weight variance for $\tanh$ at a bias variance of $0.05$ and of order $N^{-1/2}$ at zero bias: towards larger weights for a typical network, the other way for the mean over networks. On the mean-field edge of chaos a network of width $64$ is in its ordered phase, contracting the distance between two inputs, and the squared gradient, by a factor $14$ over a hundred layers. Because signals and gradients follow the same walk, one condition makes a typical network critical for both, and there the distance performs a random walk without drift: an avalanche \cite{ghavasieh2025a,ghavasieh2025b}, a run of layers in which the distance stays above its starting value, is an excursion of such a walk, with survival probability $n^{-1/2}$ [Fig.~\ref{fig:walk}(d)] and exponents $3/2$ for the duration and $4/3$ for the size \cite{sparre1954,bray2013,christensen2005} (\SM, Sec.~S6), the class measured in Refs.~\cite{ghavasieh2025a,ghavasieh2025b}.

Along the edge of chaos, small signals lose no information to noise, and the plateau set by $N/L$ agrees with the aspect ratio as the control parameter of the $1/N$ expansion \cite{hanin_nica2020,jakub2024,combette2026,pdlt,hanin2024}. ReLU, with a single critical point, must compress and transmits less information; the leaky ReLU shaped to the depth recovers part of the difference (Table~\ref{tab:mnist}). In practice ReLU networks rely on normalisation layers and skip connections, which our tests lack.

The training runs test the operating point and the first layer more than the $1/N$ correction, which leaves the final accuracy unchanged within the error and shows in the training speed and in the propagation of signals and gradients (Secs.~III\,B and C). The theory treats fully connected networks with Gaussian or orthogonal weights at initialisation; convolutional and residual layers change the recursions \cite{tamai2025}, and training moves the weights. The training tests use two small data sets, two widths and two depths. In deep, narrow networks ($L>N$) the zero-bias point of the finite-width edge of chaos compresses more than the budget allows ($\gamma=1.2$ at $L=96$, $N=64$, and $1.9$ at $N=32$), and Crinit cannot lower the compression with a non-negative bias variance; a wider network or the leaky ReLU shaped to the depth avoids this.

Wherever propagation is a product of random factors, as in recurrent networks \cite{sompolinsky1988} and models of neural activity in the brain \cite{williamsgarcia2014,beggs2003}, the mean and the typical value have critical points about one over the system size apart, and a single sample sees the typical one.

\lastheading\begin{acknowledgments}
The author declares no competing interests. The author used the language model Claude Opus 5.5 (Anthropic) to write and debug the code, to draft the article and the Supplemental Material from the author's notes and instructions, to edit the text and to check derivations; the author directed this use, revised the text and verified all of its output.
\end{acknowledgments}

\lastheading\section*{Data availability}
The data that support the findings of this article are contained in the article and its Supplemental Material \cite{sm}; the code that generates them and the Crinit package are openly available \cite{code}.

%

\clearpage\onecolumngrid\pagestyle{empty}
\newpage\thispagestyle{empty}\noindent\hspace*{-1in}\hspace*{-\oddsidemargin}\smash{\raisebox{\dimexpr-\paperheight+1in+\topmargin+\headheight+\headsep+\topskip\relax}[0pt][0pt]{\includegraphics[page=1,width=\paperwidth,height=\paperheight,keepaspectratio]{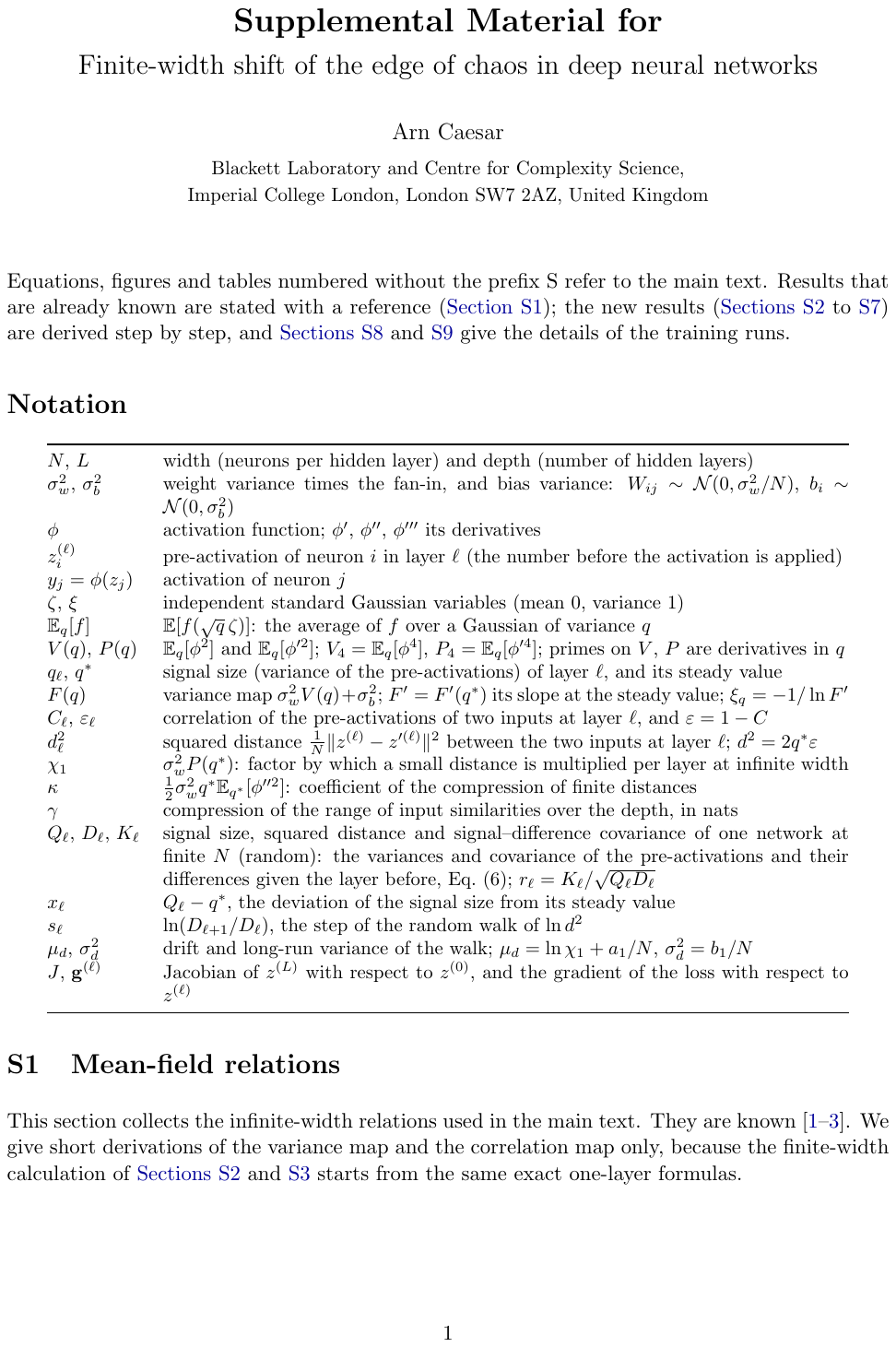}}}
\newpage\thispagestyle{empty}\noindent\hspace*{-1in}\hspace*{-\oddsidemargin}\smash{\raisebox{\dimexpr-\paperheight+1in+\topmargin+\headheight+\headsep+\topskip\relax}[0pt][0pt]{\includegraphics[page=2,width=\paperwidth,height=\paperheight,keepaspectratio]{sm.pdf}}}
\newpage\thispagestyle{empty}\noindent\hspace*{-1in}\hspace*{-\oddsidemargin}\smash{\raisebox{\dimexpr-\paperheight+1in+\topmargin+\headheight+\headsep+\topskip\relax}[0pt][0pt]{\includegraphics[page=3,width=\paperwidth,height=\paperheight,keepaspectratio]{sm.pdf}}}
\newpage\thispagestyle{empty}\noindent\hspace*{-1in}\hspace*{-\oddsidemargin}\smash{\raisebox{\dimexpr-\paperheight+1in+\topmargin+\headheight+\headsep+\topskip\relax}[0pt][0pt]{\includegraphics[page=4,width=\paperwidth,height=\paperheight,keepaspectratio]{sm.pdf}}}
\newpage\thispagestyle{empty}\noindent\hspace*{-1in}\hspace*{-\oddsidemargin}\smash{\raisebox{\dimexpr-\paperheight+1in+\topmargin+\headheight+\headsep+\topskip\relax}[0pt][0pt]{\includegraphics[page=5,width=\paperwidth,height=\paperheight,keepaspectratio]{sm.pdf}}}
\newpage\thispagestyle{empty}\noindent\hspace*{-1in}\hspace*{-\oddsidemargin}\smash{\raisebox{\dimexpr-\paperheight+1in+\topmargin+\headheight+\headsep+\topskip\relax}[0pt][0pt]{\includegraphics[page=6,width=\paperwidth,height=\paperheight,keepaspectratio]{sm.pdf}}}
\newpage\thispagestyle{empty}\noindent\hspace*{-1in}\hspace*{-\oddsidemargin}\smash{\raisebox{\dimexpr-\paperheight+1in+\topmargin+\headheight+\headsep+\topskip\relax}[0pt][0pt]{\includegraphics[page=7,width=\paperwidth,height=\paperheight,keepaspectratio]{sm.pdf}}}
\newpage\thispagestyle{empty}\noindent\hspace*{-1in}\hspace*{-\oddsidemargin}\smash{\raisebox{\dimexpr-\paperheight+1in+\topmargin+\headheight+\headsep+\topskip\relax}[0pt][0pt]{\includegraphics[page=8,width=\paperwidth,height=\paperheight,keepaspectratio]{sm.pdf}}}
\newpage\thispagestyle{empty}\noindent\hspace*{-1in}\hspace*{-\oddsidemargin}\smash{\raisebox{\dimexpr-\paperheight+1in+\topmargin+\headheight+\headsep+\topskip\relax}[0pt][0pt]{\includegraphics[page=9,width=\paperwidth,height=\paperheight,keepaspectratio]{sm.pdf}}}
\newpage\thispagestyle{empty}\noindent\hspace*{-1in}\hspace*{-\oddsidemargin}\smash{\raisebox{\dimexpr-\paperheight+1in+\topmargin+\headheight+\headsep+\topskip\relax}[0pt][0pt]{\includegraphics[page=10,width=\paperwidth,height=\paperheight,keepaspectratio]{sm.pdf}}}
\newpage\thispagestyle{empty}\noindent\hspace*{-1in}\hspace*{-\oddsidemargin}\smash{\raisebox{\dimexpr-\paperheight+1in+\topmargin+\headheight+\headsep+\topskip\relax}[0pt][0pt]{\includegraphics[page=11,width=\paperwidth,height=\paperheight,keepaspectratio]{sm.pdf}}}
\newpage\thispagestyle{empty}\noindent\hspace*{-1in}\hspace*{-\oddsidemargin}\smash{\raisebox{\dimexpr-\paperheight+1in+\topmargin+\headheight+\headsep+\topskip\relax}[0pt][0pt]{\includegraphics[page=12,width=\paperwidth,height=\paperheight,keepaspectratio]{sm.pdf}}}
\newpage\thispagestyle{empty}\noindent\hspace*{-1in}\hspace*{-\oddsidemargin}\smash{\raisebox{\dimexpr-\paperheight+1in+\topmargin+\headheight+\headsep+\topskip\relax}[0pt][0pt]{\includegraphics[page=13,width=\paperwidth,height=\paperheight,keepaspectratio]{sm.pdf}}}
\newpage\thispagestyle{empty}\noindent\hspace*{-1in}\hspace*{-\oddsidemargin}\smash{\raisebox{\dimexpr-\paperheight+1in+\topmargin+\headheight+\headsep+\topskip\relax}[0pt][0pt]{\includegraphics[page=14,width=\paperwidth,height=\paperheight,keepaspectratio]{sm.pdf}}}
\newpage\thispagestyle{empty}\noindent\hspace*{-1in}\hspace*{-\oddsidemargin}\smash{\raisebox{\dimexpr-\paperheight+1in+\topmargin+\headheight+\headsep+\topskip\relax}[0pt][0pt]{\includegraphics[page=15,width=\paperwidth,height=\paperheight,keepaspectratio]{sm.pdf}}}
\newpage\thispagestyle{empty}\noindent\hspace*{-1in}\hspace*{-\oddsidemargin}\smash{\raisebox{\dimexpr-\paperheight+1in+\topmargin+\headheight+\headsep+\topskip\relax}[0pt][0pt]{\includegraphics[page=16,width=\paperwidth,height=\paperheight,keepaspectratio]{sm.pdf}}}
\newpage\thispagestyle{empty}\noindent\hspace*{-1in}\hspace*{-\oddsidemargin}\smash{\raisebox{\dimexpr-\paperheight+1in+\topmargin+\headheight+\headsep+\topskip\relax}[0pt][0pt]{\includegraphics[page=17,width=\paperwidth,height=\paperheight,keepaspectratio]{sm.pdf}}}
\newpage\thispagestyle{empty}\noindent\hspace*{-1in}\hspace*{-\oddsidemargin}\smash{\raisebox{\dimexpr-\paperheight+1in+\topmargin+\headheight+\headsep+\topskip\relax}[0pt][0pt]{\includegraphics[page=18,width=\paperwidth,height=\paperheight,keepaspectratio]{sm.pdf}}}
\newpage\thispagestyle{empty}\noindent\hspace*{-1in}\hspace*{-\oddsidemargin}\smash{\raisebox{\dimexpr-\paperheight+1in+\topmargin+\headheight+\headsep+\topskip\relax}[0pt][0pt]{\includegraphics[page=19,width=\paperwidth,height=\paperheight,keepaspectratio]{sm.pdf}}}
\newpage\thispagestyle{empty}\noindent\hspace*{-1in}\hspace*{-\oddsidemargin}\smash{\raisebox{\dimexpr-\paperheight+1in+\topmargin+\headheight+\headsep+\topskip\relax}[0pt][0pt]{\includegraphics[page=20,width=\paperwidth,height=\paperheight,keepaspectratio]{sm.pdf}}}
\newpage\thispagestyle{empty}\noindent\hspace*{-1in}\hspace*{-\oddsidemargin}\smash{\raisebox{\dimexpr-\paperheight+1in+\topmargin+\headheight+\headsep+\topskip\relax}[0pt][0pt]{\includegraphics[page=21,width=\paperwidth,height=\paperheight,keepaspectratio]{sm.pdf}}}
\newpage\thispagestyle{empty}\noindent\hspace*{-1in}\hspace*{-\oddsidemargin}\smash{\raisebox{\dimexpr-\paperheight+1in+\topmargin+\headheight+\headsep+\topskip\relax}[0pt][0pt]{\includegraphics[page=22,width=\paperwidth,height=\paperheight,keepaspectratio]{sm.pdf}}}
\newpage\thispagestyle{empty}\noindent\hspace*{-1in}\hspace*{-\oddsidemargin}\smash{\raisebox{\dimexpr-\paperheight+1in+\topmargin+\headheight+\headsep+\topskip\relax}[0pt][0pt]{\includegraphics[page=23,width=\paperwidth,height=\paperheight,keepaspectratio]{sm.pdf}}}
\newpage\thispagestyle{empty}\noindent\hspace*{-1in}\hspace*{-\oddsidemargin}\smash{\raisebox{\dimexpr-\paperheight+1in+\topmargin+\headheight+\headsep+\topskip\relax}[0pt][0pt]{\includegraphics[page=24,width=\paperwidth,height=\paperheight,keepaspectratio]{sm.pdf}}}
\newpage\thispagestyle{empty}\noindent\hspace*{-1in}\hspace*{-\oddsidemargin}\smash{\raisebox{\dimexpr-\paperheight+1in+\topmargin+\headheight+\headsep+\topskip\relax}[0pt][0pt]{\includegraphics[page=25,width=\paperwidth,height=\paperheight,keepaspectratio]{sm.pdf}}}
\newpage\thispagestyle{empty}\noindent\hspace*{-1in}\hspace*{-\oddsidemargin}\smash{\raisebox{\dimexpr-\paperheight+1in+\topmargin+\headheight+\headsep+\topskip\relax}[0pt][0pt]{\includegraphics[page=26,width=\paperwidth,height=\paperheight,keepaspectratio]{sm.pdf}}}
\newpage\thispagestyle{empty}\noindent\hspace*{-1in}\hspace*{-\oddsidemargin}\smash{\raisebox{\dimexpr-\paperheight+1in+\topmargin+\headheight+\headsep+\topskip\relax}[0pt][0pt]{\includegraphics[page=27,width=\paperwidth,height=\paperheight,keepaspectratio]{sm.pdf}}}
\newpage\thispagestyle{empty}\noindent\hspace*{-1in}\hspace*{-\oddsidemargin}\smash{\raisebox{\dimexpr-\paperheight+1in+\topmargin+\headheight+\headsep+\topskip\relax}[0pt][0pt]{\includegraphics[page=28,width=\paperwidth,height=\paperheight,keepaspectratio]{sm.pdf}}}
\end{document}